\documentstyle[epsfig]{elsart}

\begin{document}

\begin{frontmatter}
  
  \title {\bf Muon Decay to One Loop Order in the Left-Right Symmetric
    Model}
  
  \author{M. Czakon}
  
  \address{Institut f\"ur Theoretische Physik, Universit\"at Karlsruhe, \\
    D-76128 Karlsruhe, Germany}
  
  \author{J. Gluza} and \author{J. Hejczyk}
  
  \address{Department of Field Theory and Particle Physics,
    Institute of Physics, \\
    University of Silesia, Uniwersytecka 4, PL-40-007 Katowice,
    Poland}

\begin{abstract}

  One loop corrections to the muon decay are studied in a popular and
  self-consistent version of the Left-Right symmetric model. It is
  shown quantitatively, that the corrections do not split into those
  that come from the Standard Model sector, and some decoupling
  terms. For a heavy Spontaneous Symmetry Breaking (SSB) scale
  of the order of a least 1 TeV, the contributions from the top quark
  have a logarithmic behaviour and there is a strong quadratic
  dependence on the heavy Higgs scalar masses. The dependence on the
  light Higgs boson mass is small. The heavy neutrinos are shown to
  play an important role, although secondary in comparison with the
  heavy scalar particles as long as the heavy neutrinos' Majorana
  Yukawa coupling matrix $h_M$ obeys unitarity bounds.
    
\end{abstract}

\end{frontmatter}

\section{Introduction}

Embedding  the Standard Model (SM) into a larger gauge group increases
the number of degrees of freedom.  For the Left-Right Symmetric Models
(LRSM) based on the $SU(2)_L \otimes SU(2)_R \otimes U(1)_{B-L}$ gauge
group \cite{pati2,georgi} these are connected with new fields  and
interactions.  The model is complex  with extra particles of different
types.  New neutral leptons, charged and neutral gauge bosons, neutral
and charged Higgs scalars appear. There are many different versions of
the LR models  with equal or different left and right gauge couplings
$g_{L,R}$, and specific Higgs sector representations.  This robust
structure is a chalenge and a good  theoretical laboratory for testing
many phenomena beyond the SM.
 
The purpose of the present work is to study numerically one loop
corrections to muon decay which come from the extended gauge sector of
the LRSM.  Since the history of the LRSM is already quite long, there
have been some interesting attempts to study radiative corrections
within its framework \cite{box,fre,rizz}. To our knowledge however,
there has never been a complete calculation performed. We start our
systematic one loop level study of the model from a low-energy muon
decay  calculation.  The subject has already been explored
qualitatively in \cite{npb}.  The main result of this paper was to
show that the quadratic top mass dependence of the oblique corrections
to $\Delta r$ is lost. In the SM these corrections come from
constraints imposed by the SSB sector on the Weinberg angle
counter-term. Here similar constraints connect the counter-term with
the heavy SSB scale. As a result the top quark is effectively
massless. By the same the SM one loop corrections do not constitute a
subset of the full contributions. Therefore in general, it is not true
that one can properly fit New Physics Models (NPM) by taking one loop
SM corrections modified with tree level NPM couplings. These issues
are further explored in \cite{epj}.

The question which we wish to answer is the following: can we or can
we not accommodate the present experimental life-time of the muon
within a model that has a minimal Higgs sector structure supported by
phenomenology and the smallest possible number of unknown free
parameters? It is common wisdom that when there are many free
parameters any data can be fitted.  We show however, that this induces
a strong correlation between the heavy parameters. In fact a full
decoupling is not observed, and if the additional masses tend to
infinity independently, a huge correction results, which is
incompatible with data.

We first discuss assumptions on model's structure and parameters. The
renormalization scheme is then introduced and the corrections to muon
decay enumerated. Numerical estimates follow with a study of the
dependence on the heavy masses and the heavy symmetry breaking
scale. Conclusions together with an outlook close the paper. An
appendix gathers our notational conventions and main components of the
model.

\section{Structure and Parameters of the Model}

\label{structure}

As noticed in the Introduction, there are many versions of LRSM.  A
complete analysis at the one loop level requires the model to be
fixed. We choose to explore  the most popular version of the model
with a Higgs representation with a bidoublet $\Phi$ and two (left and
right) triplets $\Delta_{L,R}$ \cite{class}.  We also assume that the
VEV of the left-handed triplet $\Delta_{L}$ vanishes,
$\langle\Delta_{L}\rangle =0$ and the CP symmetry can be violated  by
complex phases in the quark and lepton mixing matrices. Left and right
gauge couplings are chosen to be equal, $g_L=g_R$.  We call this
model, the Minimal Left-Right Symmetric Model (MLRSM). The necessary
definitions can be found in the Appendix (for details, see
\cite{npb,class,gl92,ann}).

We also take advantage of several approximations which come from
phenomenological studies.

\begin{enumerate}

\item \underline{Mixing of Fermions}

As usual in one loop analyses, we neglect quark mixings.  The case of
neutrino mixings deserves however additional comments.  The effective
light neutrino mass matrix $M=-M_DM_R^{-1}M_D^T$ (Eq.~\ref{lek}) yields three light
Majorana neutrinos which are predominantly composed of the usual
``active'' neutrinos $\nu_L$ with a very small $O(1/m_N)$ admixture of
``sterile'' neutrinos $\nu_L^c=\nu_R$. The diagonalization of $M_R$
(Eq.~\ref{cie}) produces 3 heavy Majorana neutrinos which are mainly
composed of $\nu_R$. In order to  get light neutrino masses  at the eV
scale (as concluded from experimental data \cite{bar,dm}) and without
extra relations between $M_D$ and $M_R$ matrices, $m_N$ must be large
$m_N > 10^9(10^{13})$ GeV for the lepton (quark) see-saw mechanism.
However, we would also like to explore lower scales with $v_R$ of the
order of TeV. A crude light-heavy (LH) mixing estimate $O(1/m_N)$
would give in this case larger couplings. However, they would lead to
a problem with obtaining the light neutrino spectrum, namely, from
Eq.~\ref{lek} their masses would be much above the eV scale. A
fine-tuning of $M_D$ and/or $M_R$ parameters or additional discrete
symmetries must be applied for the full neutrino mass matrix to get
the proper light neutrino spectrum. Therefore it has been argued in
\cite{tev} that it is not natural to obtain large LH mixings for heavy
neutrinos at the TeV scale. In accordance with these arguments we
assume here that the light and heavy neutrino sectors are disconnected
(negligible mixings).  In this way, $W_1$ couples only to light
neutrinos, while $W_2$ couples to the heavy ones. $Z_1$ and $Z_2$ turn
out to couple to both of them \cite{gl92,ann}. This generates
automatically an extended flavour symmetry, where transitions outside
of a family composed from a lepton, a light and a heavy neutrino are
forbidden.

As lower limit on the heavy neutrino masses we use the direct
experimental limit from the lack of $Z \to \nu N$ decay, which is $m_N
\geq M_Z$.

\item \underline{Mixing of Charged Gauge Bosons}

In principle the model allows for mixing of charged gauge
bosons. However, experimental data analyses give the following
conservative upper bound on the mixing angle \cite{plb,pdg}
\begin{equation}
  | \xi | \leq 0.013\; \mbox{\rm rad}.
\end{equation}
The tree level contribution to $\Delta r$ coming from the mixing is
proportional to
\begin{equation}
  \sin^2 \xi \frac{M^2_{W_1}}{M^2_{W_2}}.
\end{equation}
Even if the second charged gauge boson had a mass of the order of the
SM $W$, this number would be negligible compared to the experimental
value which is of the order of 3\%.

We therefore put $\xi = 0$, which also means that $\kappa_2 =
0$. There are several advantages of this approximation. First there is
no need to renormalize the mixing of the gauge bosons. It turns also
out, that together with the previous approximation on lepton sector
mixings, there is no need to renormalize the a priori possible mixing
between the light and the heavy neutrino within a family. At last, the
$QED$ contributions to the process form a self-contained class as in
the SM (see section~\ref{corrections}).

This model has the nice feature that all the constraints on the right
handed sector come uniquely from one loop corrections. The tree level
$W_1$ exchange diagram is not sensitive to the additional gauge
structure anymore \cite{doi}.

\item \underline{Yukawa Couplings to Charged Higgs Scalars}

The approximations from the preceding two points, leave still a
possibility for muon decay through one of the charged Higgs scalars
$H_1^+$. It turns out, that the experimental data on polarized muon decay
asymmetries are compatible even with a decay through scalar currents
only. However, inverse muon decay bounds these contributions to be at
most one order of magnitude smaller than the SM left-handed current
decay. We assume in this work that these diagrams are either
negligible or require only to be included at the tree level in which
case, the space left for $\Delta r$ in Eq.~\ref{dr} would be
respectively smaller.

\end{enumerate}

\section{Renormalization in One Loop Order}

As the basic set of input parameters we choose the electromagnetic
coupling constant and the masses of the four gauge bosons, Higgs
scalars and fermions. It turns out that as long as no corrections need
to be included to tree level Higgs scalar exchange diagrams, the
on-shell conditions of gauge bosons only suffice to fix all of the
necessary counter-terms. Moreover, decoupling effects should be
automatically included.

In the present approximation, where we neglect charged gauge boson
mixing, only the Weinberg angle requires renormalization. We recall
here its counter-term \cite{npb}
\begin{eqnarray}
  \delta s_W^2 &=& 2 c_W^2
  \frac{(\delta M_{Z_2}^2+\delta M_{Z_1}^2)-(\delta M_{W_2}^2+\delta M_{W_1}^2)}
  {(M_{Z_2}^2+M_{Z_1}^2)-(M_{W_2}^2+M_{W_1}^2)}
  \nonumber \\ & & \nonumber \\ & &
  +\frac{1}{2} \frac{(M_{W_2}^2+M_{W_1}^2)(\delta M_{Z_2}^2+\delta M_{Z_1}^2)
    +(M_{Z_2}^2+M_{Z_1}^2)(\delta M_{W_2}^2+\delta M_{W_1}^2)}
  {\left((M_{Z_2}^2+M_{Z_1}^2)-(M_{W_2}^2+M_{W_1}^2)\right)^2}
  \nonumber \\ & & \nonumber \\ & &
  -\frac{1}{2} \frac{(2 M_{Z_1}^2+M_{Z_2}^2)\delta M_{Z_1}^2
    +(2 M_{Z_2}^2+M_{Z_1}^2)\delta M_{Z_2}^2}
  {\left((M_{Z_2}^2+M_{Z_1}^2)-(M_{W_2}^2+M_{W_1}^2)\right)^2}. 
\end{eqnarray}

As discussed in section~\ref{structure}, no fermion mixing
renormalization is needed, and the hard corrections (factorized weak
contributions) are properly included through the simple fermion
counter-terms
\begin{equation}
  \delta Z^{l,\nu}_{L,R} = \Sigma^{l, \nu}_{\gamma L,R},
\end{equation}
where we used the following decomposition of fermion self-energy
\begin{equation}
  \Sigma = \hat{p} P_L \Sigma_{\gamma L}+\hat{p} P_R \Sigma_{\gamma R}
           +P_L \Sigma_{L}+P_R \Sigma_{R}.
\end{equation}

An interesting problem is connected to charge universality and
renormalization of the electromagnetic coupling. We wish here to show
that charge universality follows simply from Ward identities and the
constructive proof gives also the correct counter-term. To one loop the
potentially problematic contributions come from diagrams involving a
heavy neutrino and the traditional approach gives a result independent
of these masses only after summation of vertex and external line
contributions \cite{npb}.

Let us start from the following relation, which comes from charge
assignments within a fermion doublet and the definition of the
physical fields
\begin{equation}
  \left(
    \begin{array}{c}
      0 \\
      0 \\
      \langle (l^0_L \overline{l^0_L} + \nu^0_L \overline{\nu^0_L})
      {B}^{0\; \mu}\rangle_{amp.} \\
    \end{array}
  \right)= (U_0^T)^{-1}  \left(
    \begin{array}{c}
      \langle (l^0_L \overline{l^0_L} + \nu^0_L \overline{\nu^0_L})
      {Z}_{1}^{\mu}\rangle_{amp.} \\
      \langle (l^0_L \overline{l^0_L} + \nu^0_L \overline{\nu^0_L})
      {Z}_{2}^{\mu}\rangle_{amp.} \\
      \langle l^0_L \overline{l^0_L} {A}^{\mu}\rangle_{amp.} \\
    \end{array}
  \right),
\end{equation}
where $U_0$ is the bare neutral sector mixing matrix
Eq.~\ref{fiz2} multiplied by the renormalization constants of the
physical fields
\begin{equation}
  U_0 = \left(
    \begin{array}{ccc}
      c^0_{W}c^0 & c^0_{W}s^0 & s^0_{W} \\
      -s^0_{W}s^0_{M}c^0 - c^0_{M}s^0 &
      -s^0_{W}s^0_{M}s^0 +c^0_{M}c^0 & c^0_{W}s^0_{M} \\
      -s^0_{W}c^0_{M}c^0 + s^0_{M}s^0 &
      - s^0_{W}c^0_{M}s^0 - s^0_{M}c^0 &  c^0_{W}c^0_{M}
    \end{array}
  \right)
  \left(
    \begin{array}{ccc}
      Z_{Z_{1}Z_{1}}^{\frac{1}{2}} & Z_{Z_{1}Z_{2}}^{\frac{1}{2}} &
      Z_{Z_{1}\gamma}^{\frac{1}{2}} \\
      Z_{Z_{2}Z_{1}}^{\frac{1}{2}} & Z_{Z_{2}Z_{2}}^{\frac{1}{2}} &
      Z_{Z_{2}\gamma}^{\frac{1}{2}} \\
      Z_{\gamma Z_{1}}^{\frac{1}{2}} & Z_{\gamma Z_{2}}^{\frac{1}{2}} &
      Z_{\gamma \gamma}^{\frac{1}{2}}
    \end{array}
  \right) 
\end{equation}
and $\langle \dots \rangle_{amp.}$ is a shorthand for amputated Green
functions. From this we obtain
\begin{equation}
  \langle l^0_L \overline{l^0_L} {A}^{\mu} \rangle_{amp.} = (U_0^T)_{33}
  \langle (l^0_L \overline{l^0_L} + \nu^0_L \overline{\nu^0_L}) {B}^{0\; \mu}
  \rangle_{amp.}.
\end{equation}
After taking the divergence of the current, we can use the $U(1)$ Ward
identity for the $B$ field and the on-shell renormalization conditions
on the fermion propagators and the electromagnetic vertex, which leads
to the following identity
\begin{equation}
e = \frac{e_0}{\sqrt{\cos{2\Theta^0_{W}}}} (U_0^T)_{33},
\end{equation}
which can be put into the following form
\begin{eqnarray}
 \frac{e_0}{e}( && Z_{\gamma\gamma}^{\frac{1}{2}} + \frac{\sin{\phi^0}
 \tan{\Theta^0_W} -\cos{\phi^0}\sqrt{\cos{2\Theta^0_{W}}}
 \tan{\Theta^0_W}}{\sqrt{\cos
 {2\Theta^0_{W}}}}Z^{\frac{1}{2}}_{Z_1\gamma} \\ && \left.-
 \frac{\cos{\phi^0} \tan{\Theta^0_W}+\sin{\phi^0}
 \sqrt{\cos{2\Theta^0_{W}}} \tan {\Theta^0_W}}{\sqrt{\cos{
 2\Theta^0_{W}}}} Z^{\frac{1}{2}}_{Z_2\gamma} \right) = 1. \nonumber
\end{eqnarray}
Since none of the above renormalization constants depends on the
initial fermion species, we have obtained the required charge
universality. At the same time we can expand this relation to first
order to yield the electromagnetic coupling renormalization counter-term
\begin{eqnarray}
\frac{\delta e}{e}=
&-&\frac{1}{2}\delta Z_{\gamma\gamma}\\  \nonumber
&-& \frac{\sin{\phi} \tan{\Theta_W} -\cos{\phi}\sqrt{\cos{2\Theta_{W}}}
  \tan{\Theta_W}}{\sqrt{\cos {2\Theta_{W}}}}Z^{\frac{1}{2}}_{Z_1\gamma}\\
&+& \frac{\cos{\phi} \tan{\Theta_W}+\sin{\phi}\sqrt{\cos{2\Theta_{W}}}
  \tan {\Theta_W}}{\sqrt{\cos{2\Theta_{W}}}}Z^{\frac{1}{2}}_{Z_2\gamma}.
\nonumber
\end{eqnarray}
We have checked by explicit calculation that the above formula gives the
same value as the usual approach. We would like to stress that to our
knowledge, such a formula for LR models has never been derived, although
similar methods have been used in SM analyses \cite{hollik}.

\section{Structure of Corrections to Muon Decay}

\label{corrections}

The muon life-time is parametrized through the Fermi coupling
constant, the mass of the muon and the $QED$ corrections to the
four-fermion interaction $\Delta q$, which are presently known up to
second order in the fine structure constant \cite{qed}
\begin{equation}
  \label{qed}
  \frac{1}{\tau_\mu} = \frac{G_F^2 m_\mu^5}{192 \pi^3} (1+\Delta q).
\end{equation}
The Fermi constant on the other hand is related to the tree level SM
coupling of the charged $W$ boson to fermions through
\begin{equation}
  \frac{G_F}{\sqrt 2} = \frac{e^2}{8 M_W^2 s_W^2} (1+\Delta r),
\end{equation}
where $\Delta r$ are higher order corrections to which we already made
reference.

There are two problems with these formulas, when moving from the SM to
other interactions. Let us first consider Eq.~\ref{qed}. It is based
on the assumption, that the basic process is described by the
four-fermion interaction, which in the charge conserving form is of a
pure $V-A$ type. In fact, as long as the interaction has only an
admixture of vector and axial currents, the $QED$ corrections are
finite and gauge invariant, hence meaningful. Notice however, that if
the process is induced also by right handed currents, then after
moving to the charge conserving form of the interaction (Fierz
transformation), there appear also scalar and tensor interactions,
which are known not to have a finite $QED$ correction. The same
problem occurs if we add charged scalar particles to the list. There
are two possibilities to remedy the situation, the first being of
course calculating by any means the process in the full model and
resign from the separation of $QED$ corrections. The second
possibility is somewhat simpler. If the tree level corrections from
right-handed and/or scalar interactions are of the same size as the
one loop corrections to the basic diagrams, we can simply ignore one
loop contributions to these additional currents and consider
Eq.~\ref{qed} as approximate and valid to one loop order only.

\begin{center}
\begin{figure}
\epsfig{file= 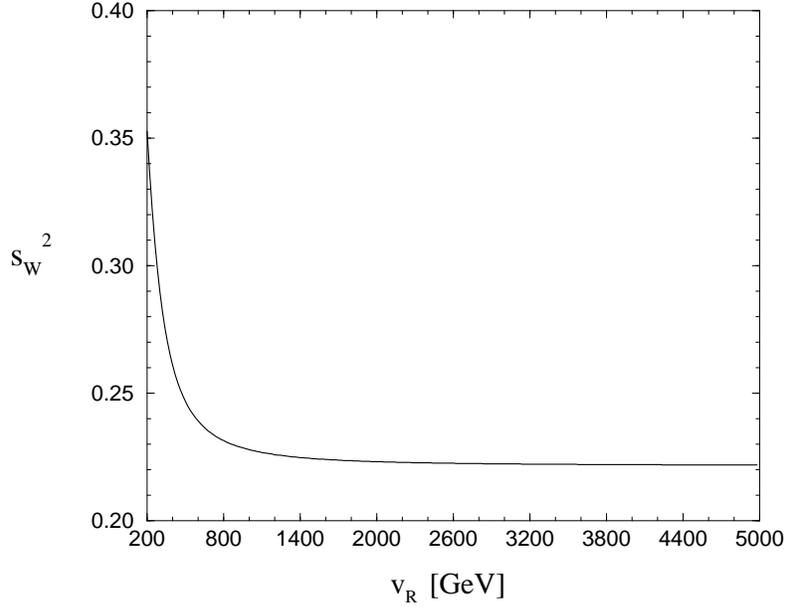, width=10cm}
\caption{$\sin^2{\Theta_W}$ as function of $v_R$ with $M_{W_1}$
  and $M_{Z_1}$ as in Eq.~\ref{val}.}
\label{sw2}
\end{figure}
\end{center}

The second problem we have to face is the fact that the tree level
coupling to the light charged gauge boson can be different from the SM
one. This concerns mainly the sine of the Weinberg angle $s_W$. In
fact this happens to be the case of the considered model, where due to
constraints if we fix the mass of the two light gauge bosons, then
$s_W$ is given by a function of the heavy SSB scale $v_R$. This
dependence is depicted in Fig.~\ref{sw2}. For small values of $v_R$,
the difference from the SM value is large. We choose here to include
the change of $s_W$ from the SM to the LRSM in $\Delta r$.

$\Delta r$ can now be obtained from the formula
\begin{eqnarray}
  \Delta r &=& \frac{(s_W^2)_{SM}}{(s_W^2)_{LRSM}} \left(
    \frac{-\Pi^T_W(0)-\delta M_W^2}{M_W^2}
    +2\frac{\delta e}{e}-\frac{\delta s_W^2}{s_W^2}+\delta_V + \delta_B \right)
  \\ \nonumber
  &-& \frac{(s_W^2)_{LRSM}-(s_W^2)_{SM}}{(s_W^2)_{LRSM}},
\end{eqnarray}
where $\delta_V$ denotes the vertex corrections, which consist of the
proper one loop vertex diagrams and the incomplete counter-term made
only of the fermion wave function renormalization constants
\begin{equation}
  \delta_V = \frac{\sqrt{2} s_W}{e} (\Lambda_{e\nu_e W} +\Lambda_{\mu
  \nu_\mu W})+\frac{1}{2}(\delta Z_L^e +\delta Z_L^{\nu_e}+\delta
  Z_L^{\mu}+\delta Z_L^{\nu_\mu}),
\end{equation}
with $\Lambda$ being the coefficient in front of the operator
$\gamma^\mu P_L$, and $\delta_B$ represents the box contributions. The
last term comes of course from the ``renormalization'' of the Weinberg
angle between the two models with
$(s_W^2)_{SM}=1-\frac{M_{W_1}^2}{M_{Z_1}^2}$ and  $(s_W^2)_{LRSM}$ as
obtained by solving Eqs.~\ref{mw},\ref{mz}.

The strong dependence on the light fermion masses in $\delta e$ is avoided as usual by a shift up to the $Z_1$ mass, and insertion of the running of the fine sturcture constant, for which we take \cite{lekk}

\begin{equation}
\Delta \alpha (M_{Z_1}) = 0.059394 \pm 0.000395
\end{equation}
The factorization of the $QED$ corrections is obtained with the Sirlin's
method \cite{sir}, which amounts to rejecting the infrared divergent box diagram
and replacing the photon vanishing mass by the $W_1$ mass in the
infrared divergent lepton wave function renormalization constants.

\section{Quantitative Results}

The evaluation of one loop corrections within the LRSM is a task of
moderate size as far as the number of diagrams is concerned. In fact
approximately 600 had to be calculated already after our simplifying
assumptions. It would not be possible to perform this work without
using an automated system. For the generation of diagrams we used the
C++ library {\bf DiaGen} \cite{diagen}, which currently contains a
topology generator, with several tools to analyse the properties of
the created objects, and a diagram generator with support for Majorana
fields. The output has then been algebraically simplified with {\bf
FORM} \cite{form}, and at last numerically evaluated with the help of
the {\bf FF} \cite{ff} based package {\bf LoopTools} \cite{looptools}.

As discussed in the previous section we parametrized the muon lifetime
corrections coming from the LRSM through $\Delta r$, which is defined
analogously as in the SM. With the present values of the coupling
constants and masses \cite{pdg}
\begin{eqnarray}
\label{val}
  && G_F = 1.16639(1)\cdot 10^{-5}\;GeV^{-2}, \;\;\;\;
  1/\alpha = 137.0359976 \pm 0.00000050, \nonumber \\
  && M_{W_1} =  80.451 \pm 0.033 \; GeV,\;\;\;\;
  M_{Z_1} =  91.1875 \pm 0.0021 \; GeV,
\end{eqnarray}
the value of $\Delta r$ with error is
\begin{equation}
\label{dr}
  \Delta r = 0.032 \pm 0.004.
\end{equation}
We depicted this experimentally allowed range by a shaded region on
the relevant figures.

\begin{center}
\begin{figure}
\epsfig{file= 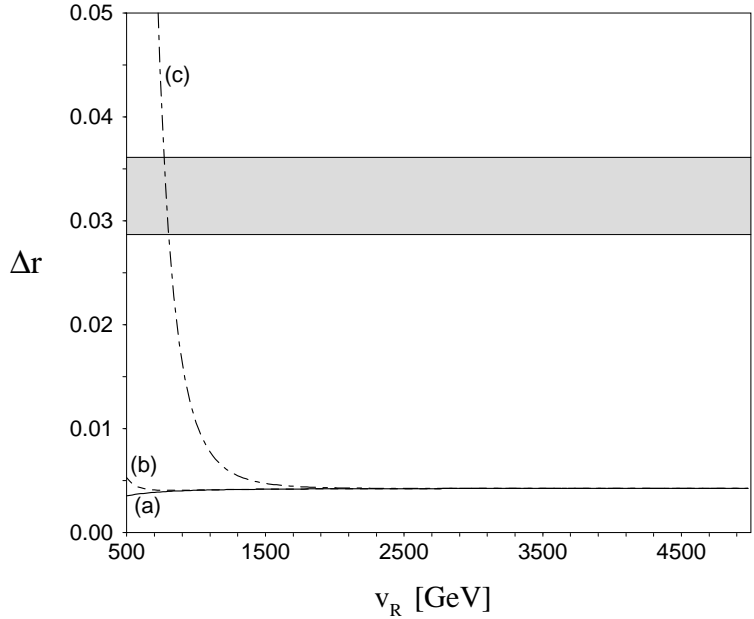, width=10cm}
\caption{Contribution of box diagrams to $\Delta r$. With increasing 
$v_R$ heavy particles decouple and the lines aim at the SM
contribution. The (a) line is for (three heavy neutrinos) $m_N=100$
GeV; (b) is for   $m_N=500$ GeV; (c)  is for $m_N=2$ TeV. Higgs
particle masses obey Eqs.~\ref{ha},\ref{hb}.  The gray area shows the
experimentally allowed values of $\Delta r$.}
\label{box}
\end{figure}
\end{center}

As noted already in a previous work \cite{npb}, we should not expect
decoupling in the sense that for large $v_R$ and large masses of the
additional particles the SM result for $\Delta r$ would be
obtained. Some type of decoupling is however observable. For example
if we take the box diagrams in the 't Hooft-Feynman gauge, then the
result tends to the SM one as depicted in Fig.~\ref{box}. It is worth
noting however the effect of taking heavy neutrinos with a low $v_R$
as for the (c) curve, where the contribution blows up. This is simply
a consequence of the fact, that the ratio of a neutrino mass and the
heavy SSB scale is proportional to the Yukawa coupling $h_M$
(Eq.~\ref{hm}) and the respective diagrams are proportional to at least
the square of these couplings. Obviously, if the Yukawa couplings
start to be larger than one then the perturbative expansion must break
down.

\begin{center}
\begin{figure}
\epsfig{file= 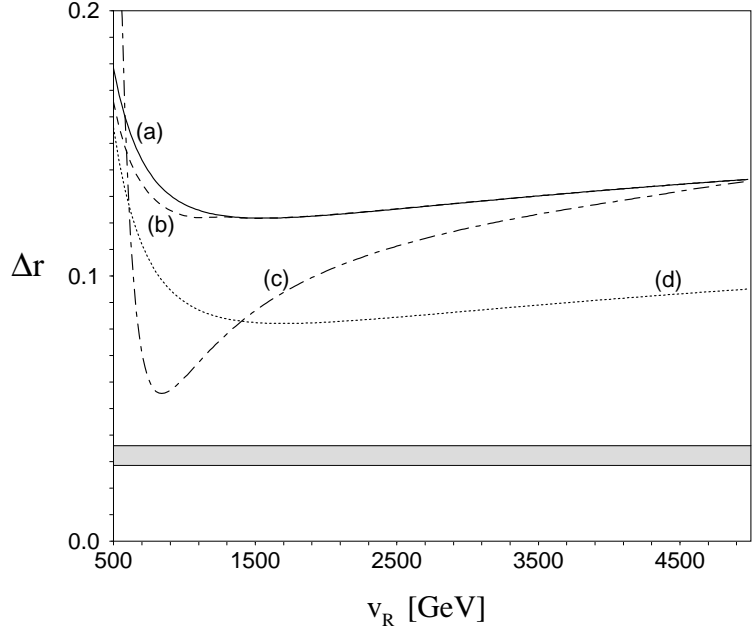, width=10cm}
\caption{
$\Delta r$ as function of $v_R$ for different heavy neutrino masses.
Higgs masses are chosen according to Eqs.~\ref{ha},\ref{hb}.  The (a)
line is for (three heavy neutrinos) $m_N=100$ GeV; (b) is for
$m_N=500$ GeV; (c)  is for $m_N=2$ TeV. Line (d) shows the results
when heavy neutrino masses follow from $h_M=1$ (see
Eq.~\ref{hm1}). The gray area shows the  experimentally allowed values
of $\Delta r$.}
\label{delta_r}
\end{figure}
\end{center}

An interesting effect is obtained, if we take the masses of the
Higgses to follow some simple pattern as in the Appendix Eqs.~\ref{ha}
and \ref{hb}. The respective $\Delta r$ is shown for several heavy
neutrino masses in Fig.~\ref{delta_r}. With growing $v_R$ the value
grows strongly away from the allowed range and these parameters must
be rejected. Although heavy neutrinos lower down $\Delta r$, we cannot
obtain a reasonable value even if their masses are at the edge of the
perturbatively range.  The line  (d) realizes this situation with the
largest possible heavy neutrino mass as a function of $v_R$
(Eq.~\ref{hm})
\begin{equation}
m_N = \sqrt{2} v_R.
\label{hm1}
\end{equation}

\begin{center}
\begin{figure}
\epsfig{file= 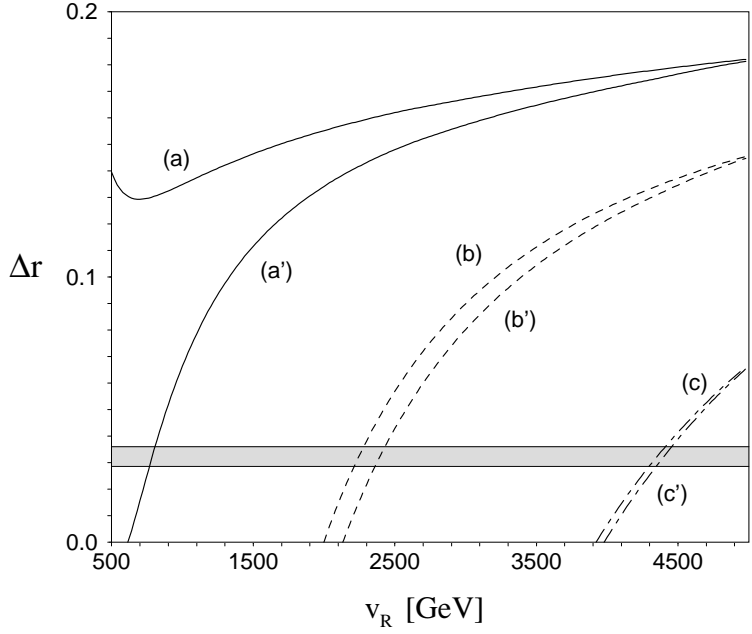, width=10cm}
\caption{$\Delta r$ as function of $v_R$. Sets with and without
primes show results for three heavy neutrino masses with $m_N=100$ GeV
and $m_N=2$ TeV respectively.  The lines describe different values of
Higgs scalar masses: (a) is for all Higgs masses  $M_{H}=1$ TeV; (b)
is for  $M_{H}=5$ TeV; (c) is for  $M_{H}=10$ TeV. The gray area shows
the experimentally allowed values of $\Delta r$.}
\label{mn}
\end{figure}
\end{center}

\begin{center}
\begin{figure}
\epsfig{file= 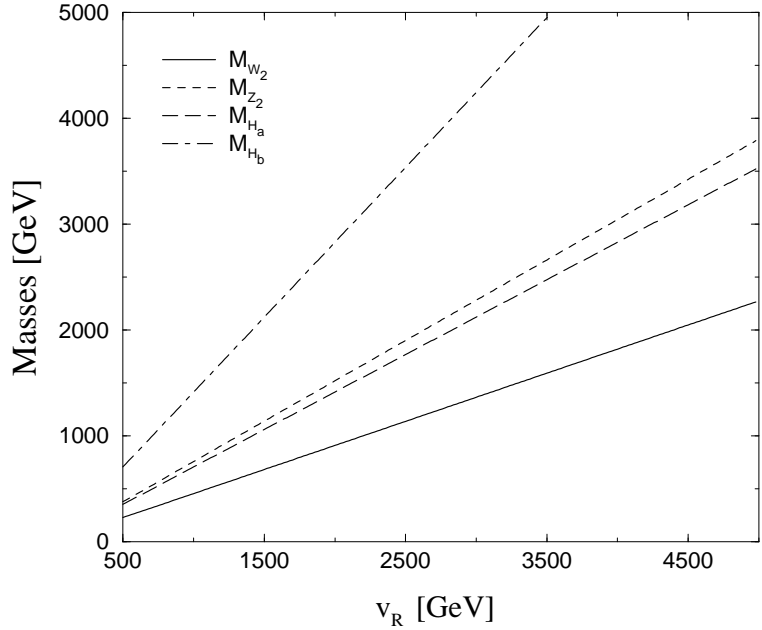, width=10cm}
\caption{Masses of additional gauge bosons and two sets of Higgs
scalar particles Eqs.~\ref{ha},\ref{hb} as function of the $v_R$ scale.}
\label{masy}
\end{figure}
\end{center}

If we now assume for simplicity that all of the Higgs scalar masses
are equal, apart from the SM Higgs boson, then we obtain the strong
dependence as depicted in Fig.~\ref{mn}. If all the scalars are
approximately two times heavier than $v_R$ (for large Higgs masses),
the experimental value for the muon decay life-time can be
accommodated.  Let us note at this point that large Higgs masses, at
least of the order of a few TeV  are needed because of FCNC
\cite{fre}.   It is obvious from Fig.~\ref{mn} that Higgs scalars,
heavy neutrinos and  additional gauge boson masses are very much
fine-tuned to be within the SM gray area, e.g. the line (a') with
$m_H=1$ TeV and $m_N=100$ GeV gives  $v_R \simeq 800$ GeV, which fixes
$M_{W_2}$ and $M_{Z_2}$ to the values as in Fig.~\ref{masy}.

\begin{center}
\begin{figure}
\epsfig{file= 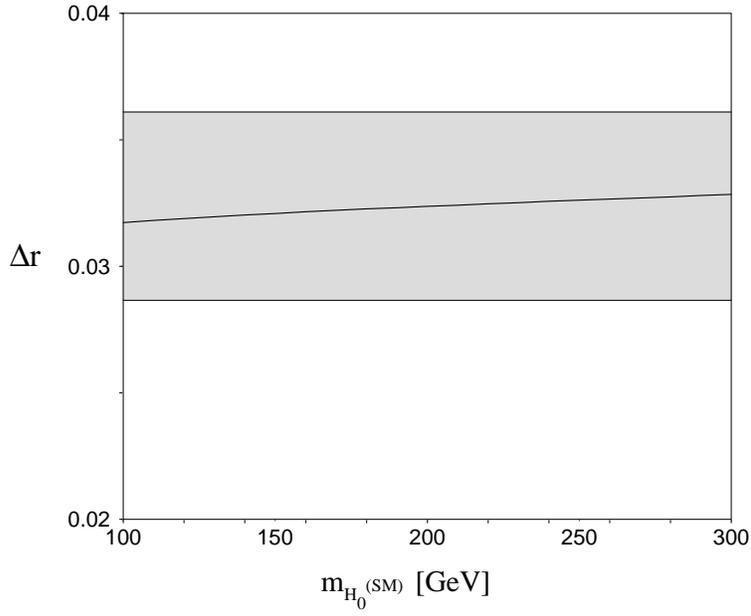, width=10cm}
\caption{$\Delta r$ as function of the lightest Higgs scalar mass 
$M_{H_0^0}$.The gray area shows the experimentally allowed values of
$\Delta r$ and the heavy particle spectrum is chosen to fit
approximately to this region, namely, $v_R=2390$ GeV, $m_N=2$ TeV,
$m_H=5$ TeV.}
\label{hsm}
\end{figure}
\end{center}

It is interesting, that for a larger SSB
breaking scale, the variation of $\Delta r$ with the SM Higgs scalar
mass is negligible. This is shown in Fig.~\ref{hsm} for  $v_R=2390$ GeV
 and neutrino and heavy scalar masses chosen to fit
the experimental value.

\begin{center}
\begin{figure}
\epsfig{file= 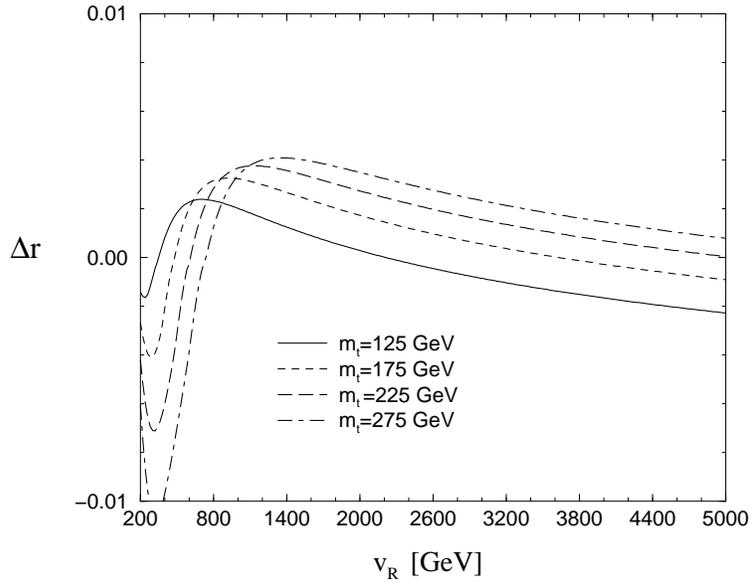, width=10cm}
\caption{The contribution of the third quark family to $\Delta r$ as 
function of $v_R$ for different top quark masses.}
\label{top}
\end{figure}
\end{center}

At last let us comment on the dependence on the top quark mass. We
show the contributions of the third quark family in
Fig.~\ref{top}. The values of the top mass spread over a vary large
range to show the behaviour of the correction. As already forseen in
\cite{npb}, for low values of $v_R$ the variation is described by a
negative quadratic function. However for a $v_R$ as low as $1\;\;TeV$,
only a positive logarithmic contribution is visible.  Notice also,
that even in the low $v_R$ range, the top mass squared enters with a
smaller coefficient than in the SM.

\section{Conclusions}

In this paper we have studied the full one loop corrections to the
muon decay in a self consistent Left-Right symmetric model. We have
shown quantitatively that the contributions have a different structure
from the SM ones and that they cannot be separated into these and some
corrections that would vanish with $v_R$. Moreover, we have shown that
the muon decay alone already puts some stringent restrictions on the
different heavy particle masses with respect to the heavy SSB breaking
scale.

Our analysis should be extended to cover also other low-energy
experiments \cite{plb}. This should elucidate the question of the
contribution of the $H^+_1$ boson to the muon decay. It will then also
be possible to derive bounds on the extra boson masses.

At last let us note that several of the assumptions that we here took
could be raised, but it is doubtful that this would change qualitatively
the numerical results. On the other hand it would certainly make the
analysis more involved, starting from the necessity to renormalize the
$\xi$ angle, LH neutrino mixing and ending with problems with the
$QED$ contributions.

\section*{Acknowledgements}

M. C. would like to thank the Alexander von Humboldt foundation for
fellowship.  This work was partly supported by the Polish Committee
for Scientific Research under Grants No. 2P03B04919 and 2P03B05418.

\section{Appendix}

In this appendix we gather our definitions and give a short account of the
particle content of the model.

\subsection{Higgs sector}

The Higgs sector contains one bidoublet and two triplets
\begin{equation}
  \Phi = \left( \matrix{ \phi_1^0 & \phi_1^+ \cr \phi_2^- & \phi_2^0 \cr}
  \right),\;\;\;\; 
  \Delta_{L,R}=  \left( \matrix{ \delta_{L,R}^+/\sqrt{2} & \delta_{L,R}^{++}
      \cr \delta_{L,R}^0 & -\delta_{L,R}^+/\sqrt{2} \cr } \right),
  \label{mul}
\end{equation}
with the allowed Vacuum Expectation Values (VEV)
\begin{equation}
  \langle \Phi \rangle = \frac{1}{\sqrt{2}} \left( \matrix{\kappa_1 & 0 \cr
      0 & \kappa_2} \right),\;
  \langle \Delta_{L,R} \rangle = \frac{1}{\sqrt{2}} \left( \matrix{0 & 0 \cr
      v_{L,R} & 0} \right),
\end{equation}
of which $\kappa_2$ and $v_L$ are assumed to vanish.

The full potential has been studied in \cite{class,gl92,ann}. Here we
only recall the physical spectrum of the particles, which consists of
\begin{itemize}
\item[(i)] four neutral scalars with $J^{PC}=0^{++}\;\;
\left( H^0_i\;\;i=0,1,2,3\right),$
\item[(ii)] two neutral pseudoscalars with $J^{PC}=0^{+-}\;\;
\left( A^0_i\;\;i=1,2\right),$
\item[(iii)] two singly charged bosons
$\left( H^{\pm}_i\;\;i=1,2\right),$ and
\item[(iv)] two doubly charged Higgs particles
$\left( \delta_L^{\pm \pm},\;
\delta_R^{\pm \pm}\right)$.
\end{itemize}

If $v_R \gg \kappa_1$ and all the parameters of the potential which
enter the Higgs masses are taken to be 1 (these are combinations of
$\mu_{1,2},\lambda_{1,...,6},\rho_{1,...,4}$ defined in \cite{class}),
then neglecting terms proportional to the VEV of the SM, the masses
satisfy the relations
\begin{eqnarray}
  M_{H_a} & \equiv & M_{H_1^0}=M_{H_3^0}=M_{A_1^0}=M_{A_2^0}
  =M_{H_1^+}=M_{H_2^+}=M_{\delta_L^{++}} = v_R/\sqrt{2}, \nonumber \\
  &&  \label{ha} \\
  M_{H_b} & \equiv  & M_{H_2^0}=M_{\delta_R^{++}}=\sqrt{2} v_R,  \label{hb} \\
  M_{H_0^0} & = & \sqrt{2} \kappa_1 .
\end{eqnarray}

\subsection{Gauge boson sector}

Gauge boson masses are generated by the  following mass terms (with
the assumption of equal couplings for the two $SU(2)$ groups)
\begin{eqnarray}
  \label{lm}
  L_M &=&
  \left(
    W_L^{+\mu},\;W_R^{+\mu} \right) M_{Charged}^2 \left(
    \begin{array}{c}
      W^-_{L\mu} \\
      W^-_{R\mu}
    \end{array} \right)+h.c. \\
  &+& \frac{1}{2} \left(W^\mu_{3L},\;W^\mu_{3R},\;B^\mu\right)
  M_{Neutral}^2\left(
    \begin{array}{c}
      W_{3L\mu} \\
      W_{3R\mu} \\
      B_\mu
    \end{array}
  \right), \nonumber
\end{eqnarray}
with
\begin{eqnarray}
  M_{Charged}^2&=&\frac{g^2}{4} \left( \matrix{ \kappa_+^2 &
      -2\kappa_1\kappa_2 \cr -2\kappa_1\kappa_2 & \kappa_+^2+2v_R^2 } \right) ,
\end{eqnarray}
and
\begin{eqnarray}
  M_{Neutral}^2&=&\frac{1}{2}\left( \matrix{ \frac{g^2}{2} \kappa_+^2 &
      -\frac{g^2}{2}\kappa_+^2 & 0 \cr -\frac{g^2}{2}\kappa_+^2 &
      \frac{g^2}{2}(\kappa_+^2+4v_R^2) & -2gg'v_R^2 \cr 0 & -2gg'v_R^2 &
      2g'^2v_R^2 } \right),
\end{eqnarray}
where $\kappa_+ =\sqrt{\kappa_1^2 +\kappa_2^2}$. The masses of the
physical gauge bosons are then given by
\begin{eqnarray}
  M^2_{W_{1,2}} &=& \frac{g^2}{4}
  \left[\kappa^2_+ +v_R^2 \mp \sqrt{v_R^4+4 \kappa^2_1\kappa^2_2}
  \right], \label{mw} \\
  M^2_{Z_{1,2}} &=& \frac{1}{4}\left\{\left[g^2 \kappa^2_+ +2v_R^2
      \left(g^2+g'^2\right)\right] \right.  \label{mz} \\
  &\mp& \left. \sqrt{\left[g^2 \kappa^2_+ +2v_R^2
          \left(g^2+g'^2\right)\right]^2-4g^2\left(g^2+2g'^2\right)\kappa^2_
        + v_R^2 }\right\}. \nonumber 
\end{eqnarray}
The symmetric mass matrices are diagonalized by the orthogonal
transformations
\begin{equation}
  \label{fiz1}
  \left(
    \begin{array}{c}
      W^{\pm}_{L} \\
      W^{\pm}_{R}
    \end{array}
  \right)=\left(
    \begin{array}{cc}
      cos\xi & sin\xi \\
      -sin\xi & cos\xi
    \end{array}
\right)=\left(
  \begin{array}{c}
    W^{\pm}_{1} \\
    W^{\pm}_{2}
  \end{array}
\right),
\end{equation}
and
\begin{equation}
  \label{fiz2}
  \left(
    \begin{array}{c}
      W_{3L} \\
      W_{3R} \\
      B
    \end{array}
  \right)=\left(
    \begin{array}{ccc}
      c_{W}c & c_{W}s & s_{W} \\
      -s_{W}s_{M}c - c_{M}s & -s_{W}s_{M}s +c_{M}c & c_{W}s_{M} \\
      -s_{W}c_{M}c + s_{M}s & - s_{W}c_{M}s - s_{M}c &  c_{W}c_{M}
    \end{array}
  \right)\left(
    \begin{array}{c}
      Z_{1} \\
      Z_{2} \\
      A
    \end{array}
  \right)
\end{equation}
where
\begin{eqnarray*}
  g &=& \frac{e}{sin{\Theta_W}},\;\;g^{\prime}=
  \frac{e}{\sqrt{\cos{2\Theta_W}}},\;\;
  c_W=\cos \Theta_W,\;\;s_W=\sin {\Theta_W}, \\
  c_M &=& \frac{\sqrt{\cos{2\Theta_W}}}{\cos{\Theta_W}},\;\;
  s_M=tg\Theta_W,\;\; s=\sin{\phi},\;\;c=\cos{\phi}.
\end{eqnarray*}
The mixing angles are given by
\begin{eqnarray}
  \tan{2\xi}=-\frac{2\kappa_1 \kappa_2}{v_R^2},\;\;\;
  \sin{2\phi}=-\frac{g^2 \kappa^2_+ \sqrt{\cos{2\Theta_W}}}{2\cos^2{\Theta_W}
    \left(
      M^2_{Z_2}-M^2_{Z_1}
    \right)}. \label{xim}
\end{eqnarray}

Fig.~\ref{masy} sums up the  mass dependence of the additional gauge
bosons and two sets of Higgs scalar particles Eqs.~\ref{ha},\ref{hb}
on the $v_R$ scale.

\subsection{Neutrinos}

The MLRSM naturally contains left as well as right handed neutrino
states. We chose the following basis for these fields
\begin{equation}
  \label{states}
  n_R = \left( \begin{array}{c} \nu^c_R \\ \nu_R \end{array} \right),
  \;\;\;  n_L = \left( \begin{array}{c} \nu_L \\ \nu_L^c \end{array}
  \right), \;\;\;  \nu^c_{L,R} = C\overline{\nu}^T_{L,R},
\end{equation}
where both $n_{L(R)}$ form 6-dimensional vectors.  The neutrino mass
matrix takes the form
\begin{eqnarray}
  M_\nu &=& \left( \begin{array}{cc} 0 & M_D \\ M^T_D & M_R
      \end{array} \right), \\ M_D &=&
      \frac{1}{\sqrt{2}}(h_l\kappa_1+\tilde{h_l}\kappa_2)=M_D^{\dagger},
      \\ M_R &=& \sqrt{2}h_Mv_R=M_R^T. \label{hm}
\end{eqnarray}
$h_l,\tilde{h}_l,h_M$ are the respective Yukawa coupling matrices.

$M_{\nu}$  can be diagonalized with the following  unitary
transformation
\begin{equation}
U = \left(\begin{array}{c} K^T_L \\ K_R^\dagger \end{array} \right),
\label{umix}
\end{equation}
where the $K_{L,R}$ matrices have dimension $6\times 3$. The {\it LEP}
neutrino counting results show that there must be three light active
neutrino states. This means that we must have $M_D \ll M_R$, which
after requiring ``natural'' couplings (order one),  turns into
$\kappa_{1,2} \ll v_R$.

Let us now introduce $3\times 3$ matrices $U_{Ll(h)}$, $U_{Rl(h)}$
\cite{gl92}:
\begin{eqnarray}
K_L&=&  \left( \matrix{U_{Ll}^\dagger \cr  U_{Lh}^\dagger } \right) ,
\\ K_R&=&  \left( \matrix{U_{Rl}^\dagger \cr  U_{Rh}^\dagger } \right).
\end{eqnarray}
The diagonalization equation assumes the form  ($m_{diag}$ and
$M_{diag}$ correspond to light and heavy neutrino mass matrices,
respectively):
\begin{eqnarray}
&& \left( \matrix{U_{Ll}^{\dagger} & U_{Rl}^T \cr U_{Lh}^{\dagger} &
           U_{Rh}^T} \right) \left( \matrix{0 & M_D \cr M_D^T &  M_R}
           \right) \left( \matrix{U_{Ll}^\ast & U_{Lh}^\ast \cr U_{Rl}
           & U_{Rh}} \right) = \left( \matrix{m_{diag} & 0 \cr 0 &
           M_{diag}} \right).
\end{eqnarray}
This can be written as:
\begin{eqnarray}
U_{Ll}^{\dagger} \left( - M_D  M_R^{-1}  M_D^T \right)   U_{Ll}^{\ast}
&  \simeq &  m_{diag}, \label{lek} \\ && \nonumber \\
U_{Rl} & \simeq  & -M_R^{-1}M_D^T  U_{Ll}^{\ast},  \\ &&  \nonumber
\\ U_{Rh}^T M_R U_{Rh}  & \simeq  & M_{diag}, \label{cie} \\ &&
\nonumber  \\ U_{Lh}^{\ast}  & \simeq  &
M_D^{\ast}{(M_R^{\ast})}^{-1}U_{Rh},
\end{eqnarray}
where the unitarity of $U$, and the large scale difference $M_{diag}
\gg m_{diag}$, have been used.  Two important conclusions can be drawn
from it:
\begin{itemize}
\item the matrices $U_{Ll}$ and $U_{Rh}$ are approximately unitary,
\item the elements of the non-diagonal submatrices $U_{Rl}$ and
$U_{Lh}$ are small, of order $\frac{<m_D>}{<M_R>} \leq \frac{O(1\;
GeV)}{m_N}$, where  $m_N=\langle M_{diag}\rangle$.
\end{itemize}
The symbol $\langle ...\rangle$ denotes the relevant scale of mass matrices.

Altogether we can write
\begin{equation}
U=  \left( \matrix{ K^T_L \cr K_R^{\dagger} } \right)= \left( \matrix{
 O(1) & O(1/m_N) \cr O(1/m_N) & O(1)} \right).
\label{hlmix}
\end{equation}

\newpage

\end{document}